\documentclass[prd,showpacs,floatfix,superscriptaddress,10 pt]{revtex4}
\usepackage{graphicx}
\usepackage{epsfig}
\usepackage{bm}
\usepackage[utf8]{inputenc}
\usepackage{amssymb}
\usepackage{float}
\usepackage{amsmath}
\usepackage{dcolumn}
\usepackage{braket}
\usepackage{physics}
\usepackage{multirow}
\usepackage{latexsym}
\usepackage{subfig}
\usepackage{amsthm}
\usepackage{framed}
\usepackage{cancel}
\usepackage[colorlinks]{hyperref}
\usepackage[usenames,dvipsnames]{color}
\hypersetup{
     breaklinks=true,
    pdfstartview={FitH},    
    colorlinks=true,       
    linkcolor=blue,          
    citecolor=red,        
    filecolor=magenta,      
    urlcolor=blue,           
    anchorcolor=green,      
    linktocpage=true
}

\makeatletter
\newcommand*{\rom}[1]{\expandafter\@slowromancap\romannumeral #1@}
\makeatother
\def\doi{http://doi.org}
\def\mes{\mathrm{mes}}
\def\mic{\mathrm{mic}}
\def\inte{\mathrm{int}}

\begin{document}

\title{Axion-Like Dark Matter Detection Using Stern-Gerlach Interferometer}

\author{Milad Hajebrahimi}
\email[]{miladhgm@gmail.com}
\affiliation{Department of Physics, Isfahan University of Technology, Isfahan 84156-83111, Iran}
\author{Hassan Manshouri}
\email[]{h.manshouri@ph.iut.ac.ir}
\affiliation{Department of Physics, Isfahan University of Technology, Isfahan 84156-83111, Iran}
\affiliation{ICRANet-Isfahan, Isfahan University of Technology, Isfahan 84156-83111, Iran}
\author{Mohammad Sharifian}
\email[]{mohammadsharifian@ph.iut.ac.ir}
\affiliation{Department of Physics, Isfahan University of Technology, Isfahan 84156-83111, Iran}
\affiliation{ICRANet-Isfahan, Isfahan University of Technology, Isfahan 84156-83111, Iran}
\author{Moslem Zarei}
\email[]{m.zarei@iut.ac.ir}
\affiliation{Department of Physics, Isfahan University of Technology, Isfahan 84156-83111, Iran}
\affiliation{ICRANet-Isfahan, Isfahan University of Technology, Isfahan 84156-83111, Iran}

\date{\today}

\begin{abstract}

Quantum sensors based on the superposition of neutral atoms are promising for sensing the nature of dark matter (DM). In this study, we utilize the Stern-Gerlach (SG) interferometer configuration to seek a novel method for the detection of detect axion-like particles (ALPs). Using an SG interferometer, we create a spatial quantum superposition of neutral atoms such as $^{3}$He and $^{87}$Rb. It is shown that the interaction of ALPs with this superposition induces a relative phase between superposed quantum components. We use the quantum Boltzmann equation (QBE) to introduce a first-principles analysis that describes the temporal evolution of the sensing system. The QBE approach employs quantum field theory (QFT) to highlight the role of the quantum nature of the interactions with the quantum systems. The resulting exclusion area demonstrates that our scheme allows for the exclusion of a range of ALP mass in the range of $10^{-10}\leq m_{a}\leq 10^{2}\,\mathrm{eV}$ and ALP-atom coupling constant in the range $10^{-13}\leq g_{ae}\leq 10^{0}$.

\end{abstract}


\maketitle

\section{Introduction}

For almost a century, observations of kinematics and mass profiles of galaxy clusters \cite{Voigt2006,Zwicky2009,Ettori2013}, and the rotation curves of nearby galaxies \cite{Sofue2001,Sofue2017} revealed that about $23\%$ of the mass-energy distribution of the Universe consists of dark matter (DM), which may include baryonic or non-baryonic matter. Up to now, however, the nature of DM has been a controversial puzzle in fundamental physics. A number of experimental proposals have been presented in order to solve this puzzle \cite{Bertone2005}. The standard model (SM) of particle physics does not include such a candidate for DM. Hence, resolving this puzzle requires going beyond the SM.

Direct detection experiments \cite{Undagoitia2015}, indirect detection observations \cite{Berton2005}, and direct production at particle colliders \cite{Buchmueller2017} are the three main methods to go looking for DM. Several recent efforts have been targeted at applying ground-based and space-based quantum sensors as new detection tools  \cite{Carney2021,Carne2021,Belenchia2022}. For example, atom interferometers \cite{Abend2019,Geiger2020} play a complementary role in DM detection in comparison with prevalent directions. The most vital feature of atom interferometers is the possibility of detecting low DM masses \cite{D2022}. Recently, atom interferometers have been used for detecting ultralight DM coupled to photons or electrons. The basic notion of these works is that DM will scatter from atoms moving in the interferometer. Such scattering induces decoherence and phase shift \cite{D2022}. Therefore, DM detection in atom interferometers is performed by two main observables: visibility as the decoherence measure of the system, and phase as the path differences measure between two arms of the interferometer \cite{D2022,Riedel2013}.

A difference between classical and quantum physical systems is the concept of ``quantum coherence'' arising from the quantum superposition principle, i.e., the possibility of existence of a single quantum state constructed by multiple states simultaneously. The waveforms of these multiple states coherently interfere with each other. One can check that coefficients of off-diagonal components in the density matrix of a system can be used as a degree measure of superposed states in the system. Additionally, quantum decoherence is the loss of coherence in a system due to interaction with the environment.

Recently, achieving quantum superposition of distinct localized states over comparatively large spatial separation to the order of $\mathrm{\mu m}$ has become an interesting topic \cite{Kamp2020}. Such superpositions can extend the trueness of the quantum world \cite{Leggett2002,Arndt2014}. One way to achieve these superposition states is to employ the Stern-Gerlach (SG) interferometer \cite{Marshman2022}.

The simplest interesting quantum systems, two-level (-state) systems, can exist in any superposition of two independent quantum states. A two-level system is used by the Unruh-DeWitt (UD) \cite{Unruh1976,DeWitt1979,Kollas2020,Foo2020,Kollas2022} particle detector, and also the SG interferometer. The UD setup has two quantum states with a given energy gap (frequency). In a recent work \cite{Kanno2022}, quantum coherence was harvested from axion DM interacting with the UD detector. The SG interferometer is based on the SG effect \cite{Gerlach1922}. In an SG experiment, a narrow beam of electrically neutral particles that act as spin $1/2$ particles is injected into an inhomogeneous magnetic field. Based on recent advances in SG interferometry, various experimental proposals have been suggested in which an inhomogeneous magnetic field is applied to effectively spin $1/2$ atoms to enable large masses to be put in spatial superposition \cite{Marshman2022,Machluf2013,Amit2019,Henkel2022,Margalit2019,Margalit2018}. A detailed investigation was also recently performed on the realization of a full-loop SG interferometer for single atoms \cite{Margalit2018}, and a new probe of quantum gravity effects was proposed \cite{Margalit2021} (for more details, see Refs. \cite{Gasbarri2021,Adelberger2022}).

For the past few decades, a theoretically new particle, known as an axion \cite{Marsh2016}, has been widely believed to be a suitable candidate for DM  (see, e.g., Refs. \cite{Chadha2022,Du2018,Graham2013,Duffy2009,Barbieri2017} and references therein). The solution suggested by Peccei and Quinn \cite{Peccei1977} to solve the strong CP problem resulted in offering the axion particle that is a pseudo-Nambu-Goldstone (pNG) boson \cite{Weinberg1978,Wilczek1987}. Furthermore, axion-like particles (ALPs) with similar features as axions are high-energy extensions of the SM axions arising from string theory (ST) \cite{Witten1984,Svrcek2006}. For ALPs, despite the QCD axions, there is no relation between mass and coupling constant. Regardless of the extremely weak coupling of axions to baryonic matter and radiation, it was proved by Sikivie \cite{Sikivie1983} that they convert into photons through the inverse Primakoff effect \cite{Primakoff1951}, and several experiments have examined this notion \cite{Zarei2022,Wuensch1989,Hagmann1990,Sikivie1985,Asztalos2001,Asztalos2010,Hoskins2011,McAllister2016,Arias2016,CAST2017,Goryachev2019}. One of the most important motivations for such theoretical and experimental investigations is that axions and ALPs with tiny mass ranges are appropriate candidates for cold DM (CDM) \cite{Dine1983,Preskill1983,Turner1983,Abbott1983,Davis1985,Davis1986,Harari1987,Battye1994,Chang1998}.

In open quantum systems (OQS), the dynamic of the density matrix of systems is studied using master equations \cite{Breuer2002}. Such equations give the time evolution of degrees of freedom of a system under the impact of its environment over the mesoscopic time scale. In this way, the quantum Boltzmann equation (QBE), as a generalization of the classical Boltzmann equation for particle occupation numbers, is a powerful tool for studying this evolution \cite{Breuer2002}. QBE for the density matrix of a system of particles is gained by taking the expectation value of the quantum number operator of the system \cite{Breuer2002}. Using QBE, Kosowsky provided a complete theoretical treatment of polarization fluctuations of cosmic microwave background (CMB) photons \cite{Kosowsky1996}. Then, several works focused on the development of QBE (see, e.g., Refs. \cite{Zarei2010,Bartolo2018,Bartolo2019,Hoseinpour2020} and references therein). Recently, QBE has been used to calculate the nuclear magnetic resonance (NMR) relaxation and decoherence times associated with fermions under the effect of a constant and an oscillating magnetic field \cite{Manshouri2021}. Additionally, the non-Markovian QBE was recently applied to investigate the damping of gravitational waves propagating in a medium consisting of decoupled ultra-relativistic neutrinos \cite{Zarei2021}.

This paper aims to study the interaction between ALPs with neutral atoms moving in an SG interferometer using quantum field theory (QFT) techniques. Considering an initial superposition state for neutral atoms, we investigate the time evolution of the density matrix of the system by QBE to numerically calculate the relative phase shift induced by ALP-atom coupling inside the SG interferometer \cite{Hatifi2020}. We consider both the forward scattering and the collision terms of the QBE, assuming the ALP scalar field exists in a coherent state. We find the exclusion regions provided by this process and compare them with other proposals for two types of neutral atoms, $^{3}$He and $^{87}$Rb.

In the rest of the paper, we use the natural unit set $c=\hbar=1$, while we know that all subsequent values of velocity components are some multiples of $c$, where $c$ is the speed of light in vacua. The paper is organized as follows. In Section \ref{eih}, we introduce the effective Hamiltonian of the process and the density matrix of the system, and we describe our scheme. In Section \ref{eef}, we study the QBE and calculate its forward scattering and collision terms related to the ALP scalar field to be in a coherent state. In Section \ref{eot}, we report the results by plotting exclusion regions for the total forward scattering and collision terms. Finally, in Section \ref{cad}, we end with conclusions.

\section{Effective interaction Hamiltonian and the density matrix of the effective two-level system}\label{eih}

In this section, we first write down the Hamiltonian density of the interacting neutral atoms and ALPs as DM in the SG interferometer configuration. To this end, we ignore the interaction between ALPs as DM and the magnetic field of the SG interferometer.

A magnetic field provides a spatial superposition for the spin particles. Their paths create a closed loop in such a way that finally recombines the different components (for more details about the full-loop SG interferometer, see Appendix \ref{ap1}). We will effectively characterize the closed loop using time- and spin-dependent momenta. We assume that neutral atoms are effectively a two-level system and can initially be split into a superposition state \cite{Ekstrom1995,Tommey2021,Frye2021,Friedrich2021}. During the movement of these neutral atoms in the interferometer, the ALPs interact with them. After recombination, the difference phase induced by this interaction contains information about the coupling and the mass of the ALPs which have occurred within the interferometer. We introduce an effective two-level system to describe the atoms in our interferometer. We assume that such a two-level system is effectively characterized by a Dirac spinor. Hence, the effective Hamiltonian density of the interaction between these neutral atoms and the ALPs as DM is expressed as
\begin{equation}\label{hd0}
\mathcal{H}_{\inte}=g\bar{\psi}\gamma^{\mu}\gamma^{5}\psi\partial_{\mu}\phi\,,
\end{equation}
in which the coupling constant of the interaction is
\begin{equation}\label{ccoi}
g=\frac{g_{af}}{m_{f}}\,,
\end{equation}
where $m_{f}$ is the mass of the two-level neutral atom system, $g_{af}$ is the dimensionless coupling constant of ALP-atom interaction, $\psi$ is the Dirac spinor with its adjoint $\bar{\psi}=\psi^{\dag}\gamma^{0}$, and $\phi(\mathbf{x},t)$ is the ALP scalar field. We also assume that the ALP scalar field is homogeneous, and hence one can set $\mu=0$ in the effective Hamiltonian density \eqref{hd0}.

To express the corresponding S-matrix element describing this process, we write down the ALP field and the spinor field in terms of the creation and the annihilation operators. The Fourier transform of the ALP field is written in terms of the creation and the annihilation operators as
\begin{equation}\label{alpssp}
\phi(\mathbf{x},t)=\phi^{-}(\mathbf{x},t)+\phi^{+}(\mathbf{x},t)\,,
\end{equation}
in which we have
\begin{equation}\label{spp}
\begin{split}
& \phi^{-}(\mathbf{x},t)=\int\frac{d^{3}\mathbf{q}}{(2\pi)^{3}\left(2q^{0}\right)}\hat{d}^{\dag}(q)
e^{\mathrm{i}(q^{0}t-\mathbf{q}\cdot\mathbf{x})}\,,\\
& \phi^{+}(\mathbf{x},t)=\int\frac{d^{3}\mathbf{q}}{(2\pi)^{3}\left(2q^{0}\right)}\hat{d}(q)
e^{-\mathrm{i}(q^{0}t-\mathbf{q}\cdot\mathbf{x})}\,,
\end{split}
\end{equation}
where $q^{0}$ is the $0$-th component of four-momentum $q$ of the ALP quantized scalar field, and $\hat{d}(q)$ and $\hat{d}^{\dag}(q)$ are the annihilation and creation operators of the ALPs, respectively, satisfying the following canonical commutation relation
\begin{equation}\label{com}
[d(q),d^{\dag}(q')]=(2\pi)^{3}\left(2q^{0}\right)\delta^{3}(\mathbf{q}-\mathbf{q}')\,.
\end{equation}
In the non-relativistic regime, $q^{0}\approx m_{a}$ where $m_{a}$ is the mass of ALPs. As mentioned, the spinor field effectively describes the two-level atom.
We can write the Fourier transform of the spinor field for neutral atoms as
\begin{equation}\label{dsplmi}
\psi(\mathbf{x},t)=\psi^{-}(\mathbf{x},t)+\psi^{+}(\mathbf{x},t)\,,
\end{equation}
in which we have
\begin{equation}\label{ds}
\begin{split}
& \psi^{-}(\mathbf{x},t)=\int\frac{d^{3}\mathbf{p}'}{(2\pi)^{3}}\sum_{r'}\bar{u}_{r'}(p')
\hat{b}_{r'}^{\dag}(p')e^{\mathrm{i}(Et-\mathbf{p}'\cdot\mathbf{x}')}\,,\\
& \psi^{+}(\mathbf{x},t)=\int\frac{d^{3}\mathbf{p}}{(2\pi)^{3}}\sum_{r}u_{r}(p)\hat{b}_{r}(p)e^{-\mathrm{i}
(Et-\mathbf{p}\cdot\mathbf{x})}\,,
\end{split}
\end{equation}
where $u_{r}$ is the free spinor solution of the Dirac equation, $p$ is the four-momentum, $\hat{b}_{r}^{\dag}(p)$ and $\hat{b}_{r}(p)$ are the creation and annihilation operators associated with spin $1/2$ systems, here neutral atoms, respectively, and the spin index $r=1,2$ counts two spin states of neutral atoms. Also, $\hat{b}_{r}^{\dag}(p)$ and $\hat{b}_{r}(p)$ obey the following canonical anti-commutation relation
\begin{equation}\label{anc}
\{b_{r}(p),b^{\dag}_{r'}(p')\}=(2\pi)^{3}\delta^{3}(\mathbf{p}-\mathbf{p}')\delta_{rr'}\,.
\end{equation}

\begin{figure}[htb]
\centering
\includegraphics[width=0.35\columnwidth]{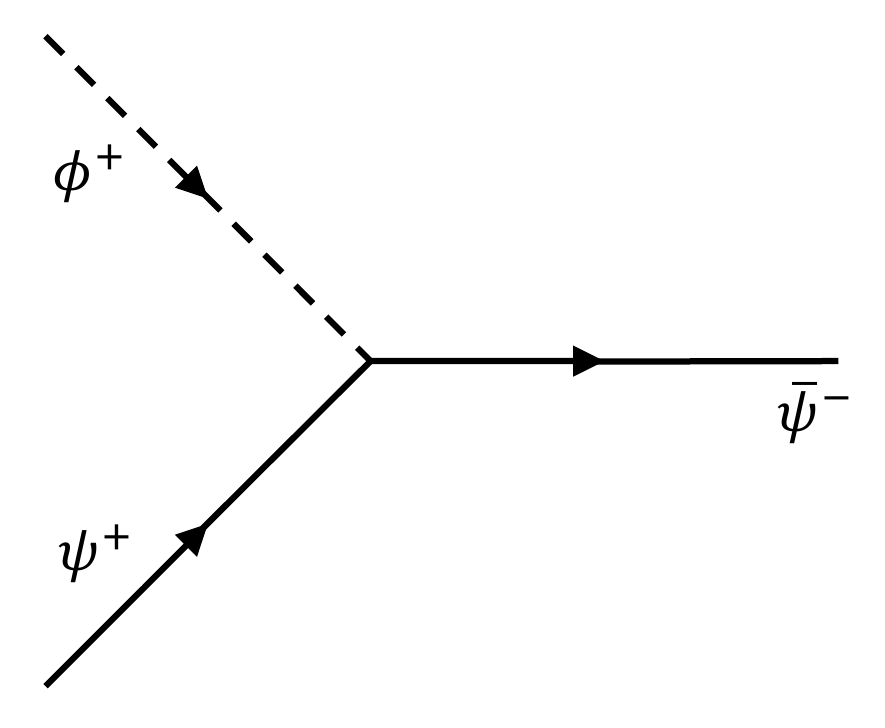}
\caption{\label{Fig1}\small{\emph{The Feynman diagram for the neutral atom-ALP interaction.}}}
\end{figure}

Fig. \eqref{Fig1} shows the scattering process. Therefore, the effective interaction Hamiltonian density describing this process can be written in the following form
\begin{equation}\label{hd}
\mathcal{H}\approx g\bar{\psi}^{-}\gamma^{0}\gamma^{5}\psi^{+}\partial_{0}\phi^{+}\,.
\end{equation}
Now, following the approach of Refs. \cite{Breuer2002,Kosowsky1996,Zarei2010,Bartolo2018,Bartolo2019,Hoseinpour2020,Manshouri2021,Zarei2021} and using \eqref{alpssp} and \eqref{dsplmi}, the effective interaction Hamiltonian $H^{0}_{\inte}(t)$ in Fourier space can be rewritten as follows
\begin{equation}\label{hhi}
H^{0}_{\inte}(t)=\int d^{3}\mathbf{x}\int\mathrm{D}\mathbf{p}\int\mathrm{D}\mathbf{p}'\int\tilde{\mathrm{D}}\mathbf{q}\,
e^{-\mathrm{i}(p^{0}-p'^{0}+q^{0})t}e^{\mathrm{i}(\mathbf{p}-\mathbf{p}'+\mathbf{q})
\cdot\mathbf{x}}(-\mathrm{i}q^{0})\mathcal{M}(\mathbf{p}\,r,\mathbf{p}'\,r')\hat{b}_{r'}^{\dag}(p')\hat{b}_{r}(p)\hat{d}(q)\,,
\end{equation}
where the superscript $0$ in $H^{0}_{\inte}(t)$ denotes that the effective interaction Hamiltonian is a functional of the free fields, and $\mathcal{M}(\mathbf{p}\,r,\mathbf{p}'\,r')=g\bar{u}_{r'}(\mathbf{p}')\gamma^{0}\gamma^{5}u_{r}
(\mathbf{p})$ is the scattering matrix element, in which $u_{r}$ is the free spinor describing the neutral atoms
\begin{equation}\label{spd}
u_{r}(\mathbf{p})=\left(
              \begin{array}{c}
                \chi_{r} \\
                \frac{\mathbf{\sigma}\cdot\mathbf{p}}{2m_{f}}\chi_{r} \\
              \end{array}
            \right)\,,
\end{equation}
where $\chi_{r}$ is the bispinor of the two-level system with spin index $r$ that generally can be read as
\begin{equation}\label{bsp}
\chi_{r}=\frac{1}{2}\left(
                      \begin{array}{c}
                        1-(-1)^{r} \\
                        1+(-1)^{r} \\
                      \end{array}
                    \right)\,.
\end{equation}
Moreover, in Eq. \eqref{hhi} we have used the following abbreviations
\begin{equation}\label{dpqpp}
\frac{d^{3}\mathbf{p}}{(2\pi)^{3}}\equiv\mathrm{D}\mathbf{p}\,,\qquad
\frac{d^{3}\mathbf{q}}{(2\pi)^{3}\left(2q^{0}\right)}\equiv\tilde{\mathrm{D}}\mathbf{q}\,.
\end{equation}

\begin{figure}[htb]
\centering
\includegraphics[width=0.6\columnwidth]{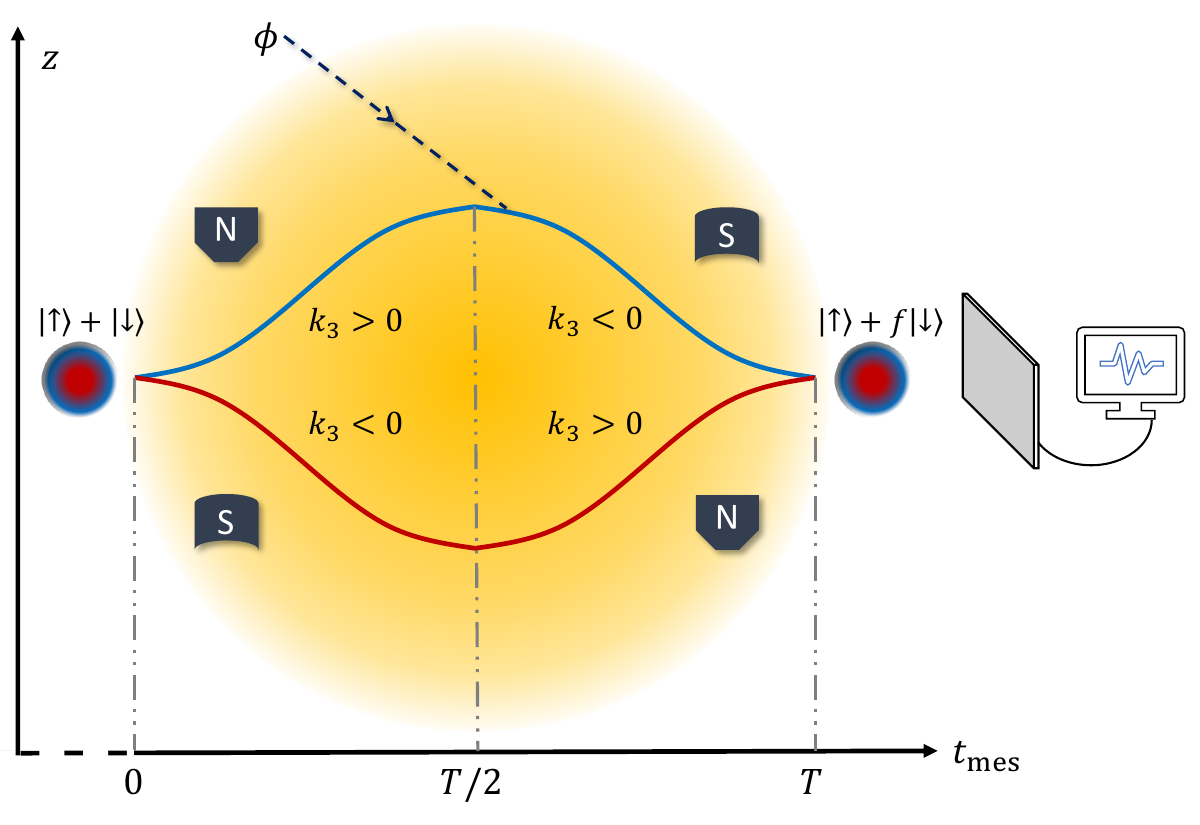}
\caption{\label{Fig2}\small{\emph{The scheme of the experimental setup over $t_{\mes}$ in which $k_{3}$ is the third component of the momentum of neutral atoms. The blue-red circle illustrates the superposition of two neutral atom states at the beginning and end of the process. One of two states, e.g., up state, proceeds in the upper loop with a blue line, and the other, the down state, moves in the lower path with a red line. Due to the effect of the magnetic field, at $t_{\mes}=0$, the superposition states of neutral atoms separate from each other so that each state (up and down) moves in one of the SG loops. Around $t_{\mes}=T/2$, the ALP DM interacts with neutral atoms. Then, at $t_{\mes}=T$, the inverse effect of the magnetic field leads to bringing the two states of neutral atoms back together (for more details, see Appendix \ref{ap1}).}}}
\end{figure}

As mentioned earlier, we provide a superposition state between spin states of neutral atoms as $(\ket{\uparrow}+\ket{\downarrow})$, in which the initial phase is considered zero. Then, the superposition state of neutral atoms interacting with ALPs in the SG interferometer transforms into $(\ket{\uparrow}+f\ket{\downarrow})$ state, where $f$ is the decoherence factor. The superposition state will be preserved during the process. On the other hand, we take into account a spin-dependent momentum for neutral atoms. Hence, two paths of the SG interferometer are dependent on the spin indices of neutral atoms. Therefore, over the mesoscopic time scale $t_{\mes}$, the sign of the momentum of neutral atoms will change between the upper and the lower paths of the SG interferometer. Consequently, the momentum of neutral atoms is time-dependent (as a similar approach, see Ref. \cite{Marshman2022}). We use the Heaviside step function $\Theta$ to write the momentum of neutral atoms. Because the $z$-axis is considered in the down-to-up direction, only the sign of the $z$-component of the momentum of neutral atoms changes during the process. This situation is illustrated in Fig. \eqref{Fig2}. This figure represents the scheme of our setup over $t_{\mes}$ on which the evolution of the mesoscopic system occurs. Hence, we can express the momentum of neutral atoms in the form of
\begin{equation}\label{fmh}
\mathbf{\mathbf{k}_{r}}(t)=\left(k^{1},k^{2},(-1)^{r}\left(-1+2\,\Theta\left[t-\frac{T}{2}\right]\right)k^{3}\right)\,,
\end{equation}
where $T$ is the duration of the whole process. Moreover, we generally have $k=(k^{0},\mathbf{k})$ where in the non-relativistic regime, we can write $k^{0}\approx m_{f}$.

One can read the density operator of a system of neutral atoms in the form of
\begin{equation}\label{dno}
\hat{\rho}=\int\mathrm{D}\mathbf{p}'\rho_{ij}(\mathbf{p}')\hat{b}_{i}^{\dag}(\mathbf{p}')\hat{b}_{j}(\mathbf{p}')\,,
\end{equation}
in which the macroscopic properties of the neutral atoms in the SG interferometer are described by $\rho_{ij}$ as the neutral atom polarization (density) matrix. Consequently, the expectation value of the neutral atom number operator $\mathcal{\hat{D}}_{ij}(\mathbf{k})=\hat{b}_{i}^{\dag}(\mathbf{k})\hat{b}_{j}(\mathbf{k})$ can be read as \cite{Kosowsky1996}
\begin{equation}\label{efn}
\left\langle\mathcal{\hat{D}}_{ij}(\mathbf{k})\right\rangle=(2\pi)^{3}\delta^{3}(0)\rho_{ji}(\mathbf{k})\,.
\end{equation}
The density matrix $\rho_{ij}(\mathbf{k})$ of a system of neutral atoms can be parameterized using Bloch vector \cite{Manshouri2021,Tolhoek1956}. Therefore, we can write
\begin{equation}\label{blv}
\rho_{ji}=\frac{1}{2}\left(\mathbb{I}+\mathbf{\sigma}\cdot\mathbf{\zeta}\right)=\frac{1}{2}\left(
                                                          \begin{array}{cc}
                                                         1+\zeta_{3} & \zeta_{1}-\mathrm{i}\zeta_{2} \\
                                                         \zeta_{1}+\mathrm{i}\zeta_{2} & 1-\zeta_{3} \\
                                                                                    \end{array}
                                                                                           \right)\,,
\end{equation}
where $\mathbb{I}$ is the $2\times 2$ identity matrix, $\mathbf{\sigma}$ is the Pauli vector made by Pauli matrices, and $\zeta_{i}$ are the components of the Bloch vector constructing the Bloch sphere. One can read $\zeta_{1}-\mathrm{i}\zeta_{2}=f$ in which the decoherence factor can be expressed as
\begin{equation}\label{decm}
f=\exp[-\epsilon+\mathrm{i}\varphi]\,,
\end{equation}
where $\epsilon$ is a dimensionless factor to measure the decoherence, and $\varphi$ is a phase measuring the path differences between two paths of the SG interferometer \cite{D2022,Riedel2013}. When $f$ differs considerably from zero, i.e., $\epsilon\lesssim 1$ (which is satisfied in the current study), the existence of the quantum superposition state can be concluded \cite{D2022}.

\section{Evolution equation for the density matrix}\label{eef}

The QBE \cite{Breuer2002,Kosowsky1996,Zarei2010,Bartolo2018,Bartolo2019,Hoseinpour2020,Manshouri2021,Zarei2021} gives the time evolution of the density matrix elements as
\begin{equation}\label{qbe1}
\begin{split}
(2\pi)^{3}\delta^{3}(0)\frac{d}{dt_{\mes}}\rho_{ji}(\mathbf{k},t_{\mes}) & =\mathrm{i}\left\langle\left[H^{0}_{\inte}(t_{\mes}),
\mathcal{\hat{D}}_{ij}(\mathbf{k},t_{\mes})\right]\right\rangle \\
& -\int_{0}^{t_{\mes}}\left\langle
\left[H^{0}_{\inte}(t_{\mes}),\left[{H^{0}}^{\dag}_{\inte}(t_{\mic}),\mathcal{\hat{D}}_{ij}(\mathbf{k},t_{\mes}
-t_{\mic})\right]\right]\right\rangle dt_{\mic}\,,
\end{split}
\end{equation}
where the interaction time scale of individual particles takes place on the microscopic time scale $t_{\mic}$. The first term on the right-hand side of Eq. \eqref{qbe1} is known as the forward scattering term, while the second is the usual collision term. The forward scattering term vanishes if the ALP scalar field is in a vacuum state. However, assuming the scalar field to be in a coherent state, this term is non-vanishing. We display operator expectation values needed here using Wick's theorem as follows \cite{Kosowsky1996}
\begin{equation}\label{expb}
\left\langle\hat{b}_{r'}^{\dag}\hat{b}_{r}\right\rangle=(2\pi)^{3}\delta^{3}(\mathbf{p}'-\mathbf{p})
\rho_{rr'}(\mathbf{p})\,,
\end{equation}
\begin{equation}\label{expd}
\left\langle\hat{d}^{\dag}(q_{2})\hat{d}(q_{1})\right\rangle=(2\pi)^{3}(2q^{0})\delta^{3}
(\mathbf{q}_{2}-\mathbf{q}_{1})\frac{1}{2}n_{a}(\mathbf{q}_{2},\mathbf{x})\,,
\end{equation}
where $n_{a}(\mathbf{q},\mathbf{x})$ is the ALP number density of momentum $\mathbf{q}$ per unit volume. One can also express the number density in terms of the velocity distribution of the ALPs as DM \cite{Evans2019}
\begin{equation}\label{vda}
n_{a}(\mathbf{v}_{a},\mathbf{x})=n_{a}(\mathbf{x})\left(\frac{4\pi}{2\sigma_{v}^{2}}\right)^{3/2}\exp \left(-\frac{\left(\mathbf{v}_{a}+\mathbf{V}^{E}\right)^{2}}{2\sigma_{v}^{2}}\right)\,,
\end{equation}
where $\mathbf{v}_{a}$ is the velocity of the ALPs, and we have $\sigma_{v}=\upsilon_{0}/\sqrt{2}$ in which $\upsilon_{0}=220\,\mathrm{km/s}$ is the circular rotation speed of the standard halo model; $\left|\mathbf{V}^{E}\right|=230\,\mathrm{km/s}$ is the velocity of the Earth with respect to the Galactic frame, and we also have $n_{a}(\mathbf{x})=\rho^{DM}/m_{a}$, with $\rho^{DM}=0.3\,\mathrm{GeV/cm^{3}}$ as the energy density of the DM near the Earth \cite{Evans2019}. Also, we have
\begin{equation}\label{expbb}
\left\langle\hat{b}_{i}^{\dag}\hat{b}_{j}\hat{b}_{k}^{\dag}\hat{b}_{l}\right\rangle\simeq
(2\pi)^{6}\delta^{3}(\mathbf{p}_{j}-\mathbf{p}_{k})\delta^{3}(\mathbf{p}_{i}-\mathbf{p}_{l})
\delta_{jk}\rho_{li}(\mathbf{p}_{l})\,.
\end{equation}
Therefore, we can find the following expectation value
\begin{equation}\label{expbdg}
\left\langle\hat{b}_{i}^{\dag}\hat{b}_{j}\hat{b}_{k}^{\dag}\hat{b}_{l}\hat{b}_{m}^{\dag}
\hat{b}_{n}\right\rangle\simeq(2\pi)^{9}\delta^{3}(\mathbf{p}_{l}-\mathbf{p}_{m})
\delta^{3}(\mathbf{p}_{i}-\mathbf{p}_{n})\delta^{3}(\mathbf{p}_{j}-\mathbf{p}_{k})
\delta_{lm}\delta_{jk}\rho_{ni}(\mathbf{p}_{n})\,.
\end{equation}

For the case where the ALP scalar field is in a coherent state, the expectation value of the annihilation operator of the ALP scalar field in the non-relativistic limit yields \cite{Kanno2022,Dvali2018}
\begin{equation}\label{expd0}
\left\langle\hat{d}(q)\right\rangle=(2\pi)^{3/2}m_{a}A\,\delta^{3}(0)\,,
\end{equation}
where the amplitude of the coherently oscillating ALP scalar field is \cite{Kanno2022,Dvali2018}
\begin{equation}\label{ampcoh}
A=2\times 10^{-6}\sqrt{\frac{\rho^{DM}}{0.3\,\mathrm{GeV/cm^{3}}}}\left(\frac{10^{-6}\,\mathrm{eV}}{m_{a}}\right)\,
\mathrm{GeV}\,.
\end{equation}

Now, from Eqs. \eqref{blv} and \eqref{qbe1}, we can find the time derivatives of Bloch vector components as the building blocks of the density matrix of neutral atoms in the following form
\begin{equation}\label{zetai}
\begin{split}
& \dot{\zeta}_{1}(\mathbf{v},t_{\mes})=\dot{\rho}_{12}(\mathbf{v},t_{\mes})+\dot{\rho}_{21}
(\mathbf{v},t_{\mes})\,,\\
& \dot{\zeta}_{2}(\mathbf{v},t_{\mes})=-\mathrm{i}\left[\dot{\rho}_{21}(\mathbf{v},t_{\mes})
-\dot{\rho}_{12}(\mathbf{v},t_{\mes})\right]\,,\\
& \dot{\zeta}_{3}(\mathbf{v},t_{\mes})=\dot{\rho}_{11}(\mathbf{v},t_{\mes})-\dot{\rho}_{22}
(\mathbf{v},t_{\mes})\,.
\end{split}
\end{equation}
where we use $\mathbf{k}=m_{f}\mathbf{v}$ in the non-relativistic regime, in which $\mathbf{v}=(v_{1},v_{2},v_{3})$ is the velocity of neutral atoms. Eq. \eqref{zetai} results in a set of coupled differential equations for the Bloch vector components. Solving these equations, one finds the relative phase shift arising from the interaction.

\subsection{Forward scattering term}\label{sfst}

The forward scattering term on the right-hand side of Eq. \eqref{qbe1} can be approximately expressed as
\begin{equation}\label{qbefs1}
\begin{split}
\mathrm{i}\left\langle\left[H^{0}_{\inte}(t_{\mes}),\mathcal{\hat{D}}_{ij}(\mathbf{k},t_{\mes})\right]
\right\rangle\simeq\left\langle\hat{d}(q)\right\rangle\left[\left\langle\hat{b}_{r'}^{\dag}
\hat{b}_{r}\hat{b}_{i}^{\dag}\hat{b}_{j}\right\rangle-\left\langle\hat{b}_{i}^{\dag}\hat{b}_{j}
\hat{b}_{r'}^{\dag}\hat{b}_{r}\right\rangle\right]\,.
\end{split}
\end{equation}
Therefore, using Eqs. \eqref{expbb} and \eqref{expd0} we get
\begin{equation}\label{qbefs2}
\begin{split}
(2\pi)^{3}\delta^{3}(0)\frac{d}{dt_{\mes}}\rho_{ji}(\mathbf{k},t_{\mes}) & =\mathrm{i}\left\langle\left[H^{0}_{\inte}(t_{\mes}),
\mathcal{\hat{D}}_{ij}(\mathbf{k},t_{\mes})\right]\right\rangle \\
& \simeq\mathrm{i}\int d^{3}\mathbf{x}\int\mathrm{D}\mathbf{p}\int\mathrm{D}\mathbf{p}'\int\tilde{\mathrm{D}}\mathbf{q}\,e^{-\mathrm{i}(p^{0}-p'^{0}+q^{0})t_{\mes}}
e^{\mathrm{i}(\mathbf{p}-\mathbf{p}'+\mathbf{q})\cdot\mathbf{x}}\mathcal{M}(\mathbf{p}\,r,\mathbf{p}'\,r',t_{\mes})\\
& \times(-\mathrm{i}q^{0})(2\pi)^{3/2}m_{a}A\,\delta^{3}(0)(2\pi)^{6}\Big(\delta^{3}(\mathbf{p}_{r}-\mathbf{k})
\delta^{3}(\mathbf{p}'_{r'}-\mathbf{k})\delta_{ri}\rho_{jr'}(\mathbf{k})\\
& -\delta^{3}(\mathbf{k}-\mathbf{p}'_{r'})\delta^{3}(\mathbf{k}-\mathbf{p}_{r})\delta_{jr'}\rho_{ri}(\mathbf{p})\Big)\,.
\end{split}
\end{equation}

After integrating over $\mathbf{x},\, \mathbf{p},\, \mathbf{p}',$ and $\mathbf{q}$ the following result is achieved
\begin{equation}\label{qbefs3}
\begin{split}
\frac{d}{dt_{\mes}}\rho_{ji}(\mathbf{k},t_{\mes})
& \simeq\frac{1}{2}\sqrt{2\pi}g m_{a}A\,e^{-\mathrm{i}m_{a}t_{\mes}} \left[u_{r'}^{\dag}(\mathbf{k},t_{\mes})\gamma^{5}u_{i}(\mathbf{k},t_{\mes})\rho_{jr'}(\mathbf{k})
-u_{j}^{\dag}(\mathbf{k},t_{\mes})\gamma^{5}u_{r}(\mathbf{k},t_{\mes})\rho_{ri}(\mathbf{k})\right]\,.
\end{split}
\end{equation}

\subsection{Collision term}\label{sct}

The integrand in the collision term on the right-hand side of Eq. \eqref{qbe1} approximately can be expressed as follows
\begin{equation}\label{intq}
\begin{split}
\left\langle
\left[H^{0}_{\inte}(t_{\mes}),\left[{H^{0}}^{\dag}_{\inte}(t_{\mic}),\mathcal{\hat{D}}_{ij}^{(f)}
(\mathbf{k},t_{\mes}-t_{\mic})\right]\right]\right\rangle & \simeq\left\langle\hat{d}_{1}(t_{\mes})\hat{d}^{\dag}_{2}(t_{\mic})\right\rangle\left[
\left\langle\hat{b}_{r'_{1}}^{\dag}\hat{b}_{r_{1}}\hat{b}_{r_{2}}^{\dag}\hat{b}_{r'_{2}}
\hat{b}_{i}^{\dag}\hat{b}_{j}\right\rangle-\left\langle\hat{b}_{r'_{1}}^{\dag}\hat{b}_{r_{1}}
\hat{b}_{i}^{\dag}\hat{b}_{j}\hat{b}_{r_{2}}^{\dag}\hat{b}_{r'_{2}}\right\rangle\right]\\
& +\left\langle\hat{d}^{\dag}_{2}(t_{mic})\hat{d}_{1}(t_{\mes})\right\rangle\left[\left\langle
\hat{b}_{i}^{\dag}\hat{b}_{j}\hat{b}_{r_{2}}^{\dag}\hat{b}_{r'_{2}}\hat{b}_{r'_{1}}^{\dag}
\hat{b}_{r_{1}}\right\rangle-\left\langle\hat{b}_{r_{2}}^{\dag}\hat{b}_{r'_{2}}
\hat{b}_{i}^{\dag}\hat{b}_{j}\hat{b}_{r'_{1}}^{\dag}\hat{b}_{r_{1}}\right\rangle\right]
\end{split}
\end{equation}
It should be noted that operators in Eq. \eqref{intq} with index ``$1$'' associated with $H^{0}_{\inte}(t_{\mes})$ are functions of $t_{\mes}$ and operators with index ``$2$'' associated with $H^{0\dagger}_{\inte}(t_{\mic})$ are functions of $t_{\mic}$.

Thus, using Eqs. \eqref{expd} and \eqref{expbdg}, one can express the QBE \eqref{qbe1} associated with collision term as
\begin{equation}\label{qbe2}
\begin{split}
& (2\pi)^{3}\delta^{3}(0)\frac{d}{dt_{\mes}}\rho_{ji}(\mathbf{k},t_{\mes})=-\int_{0}^{t_{\mes}}
\left\langle\left[H^{0}_{\inte}(t_{\mes}),\left[{H^{0}}^{\dag}_{\inte}(t_{\mic}),\mathcal{\hat{D}}_{ij}
(\mathbf{k},t_{\mes}-t_{\mic})\right]\right]\right\rangle dt_{\mic}\\
& \simeq-\int_{0}^{t_{\mes}}dt_{\mic}\int d^{3}\mathbf{x}\int d^{3}\mathbf{x}'\int\mathrm{D}\mathbf{p}_{1}\int\mathrm{D}\mathbf{p}'_{1}\int
\mathrm{D}\mathbf{p}_{2}\int\mathrm{D}\mathbf{p}'_{2}\int\tilde{\mathrm{D}}\mathbf{q}_{1}\int\tilde{\mathrm{D}}\mathbf{q}_{2}\,(q^{0}_{1})(q^{0}_{2})\\
& \times e^{-\mathrm{i}(p^{0}_{1}-p'^{0}_{1}+q^{0}_{1})t_{\mes}}e^{\mathrm{i}(\mathbf{p}_{1}-\mathbf{p}'_{1}
+\mathbf{q}_{1})\cdot\mathbf{x}}e^{\mathrm{i}(p^{0}_{2}-p'^{0}_{2}+q^{0}_{2})(t_{\mic})}
e^{-\mathrm{i}(\mathbf{p}_{2}-\mathbf{p}'_{2}+\mathbf{q}_{2})\cdot\mathbf{x}'} \mathcal{M}_{1}(\mathbf{p}_{1}\,r_{1},\mathbf{p}'_{1}\,r'_{1},\mathbf{q}_{1},t_{\mes})\mathcal{M}_{2}
(\mathbf{p}_{2}\,r_{2},\mathbf{p}'_{2}\,r'_{2},\mathbf{q}_{2},t_{\mic})\\
& \times\Big\{(2\pi)^{3}\left(2q^{0}_{1}\right)\delta^{3}(\mathbf{q}_{1}-\mathbf{q}_{2})\frac{1}{2}
n(\mathbf{q}_{1},\mathbf{x})\Big[(2\pi)^{9}\delta^{3}(\mathbf{p}'_{2}-\mathbf{k})\delta^{3}
(\mathbf{p}'_{1}-\mathbf{k})\delta^{3}(\mathbf{p}_{1}-\mathbf{p}_{2}) \delta_{r'_{2}i}\delta_{r_{1}r_{2}}\rho_{jr'_{1}}(\mathbf{k})\\
& -(2\pi)^{9}\delta^{3}(\mathbf{p}_{1}-\mathbf{k})\delta^{3}(\mathbf{p}'_{1}-\mathbf{p}'_{2}) \delta^{3}(\mathbf{k}-\mathbf{p}_{2})\delta_{r_{1}i}\delta_{jr_{2}}\rho_{r'_{2}r'_{1}}
(\mathbf{p}'_{2})\Big]+(2\pi)^{3}\left(2q^{0}_{2}\right)\delta^{3}(\mathbf{q}_{2}-\mathbf{q}_{1})
\frac{1}{2}n(\mathbf{q}_{2},\mathbf{x})\\
& \times\Big[(2\pi)^{9}\delta^{3}(\mathbf{p}'_{2}-\mathbf{p}'_{1})\delta^{3}(\mathbf{k}-\mathbf{p}_{1}) \delta^{3}(\mathbf{k}-\mathbf{p}_{2})\delta_{jr_{2}}\delta_{r'_{2}r'_{1}}\rho_{r_{1}i}(\mathbf{p}_{1}) -(2\pi)^{9}\delta^{3}(\mathbf{p}_{2}-\mathbf{p}_{1})\delta^{3}(\mathbf{p}'_{2}-\mathbf{k})\delta^{3}
(\mathbf{k}-\mathbf{p}'_{1})\delta_{jr'_{1}}\delta_{r'_{2}i}\rho_{r_{1}r_{2}}(\mathbf{p}_{1})\Big]\Big\}
\end{split}
\end{equation}
After integrating over $\mathbf{x},\, \mathbf{x}',\, \mathbf{p}_{1},\, \mathbf{p}_{2},\, \mathbf{p}'_{1},\, \mathbf{p}'_{2},$ and $\mathbf{q}_{1}$ and using Eqs. \eqref{spd}-\eqref{fmh}, one can read Eq. \eqref{qbe2} to the following form in terms of velocities of neutral atoms and ALPs
\begin{equation}\label{qbe3}
\begin{split}
& \frac{d}{dt_{\mes}}\rho_{ji}(\mathbf{v},t_{\mes})\approx g^{2}\int_{0}^{t_{\mes}}dt_{\mic}\int \mathrm{D}\mathbf{v}_{a}\frac{m_{a}n_{a}(\mathbf{x})}
{16m_{f}^{2}}\left(\frac{4\pi}{2\sigma_{v}^{2}}\right)^{3/2}\exp\left(-\frac{\left(\mathbf{v}_{a}+\mathbf{V}^{E}\right)^{2}}{2\sigma_{v}^{2}}\right)\\
& \times\Big\{-\left[\bar{\chi}_{r'}\mathbf{\sigma}\cdot(m_{f}\mathbf{v}_{r'}(t_{\mes})-\mathbf{q})\chi_{r}
+\bar{\chi}_{r'}\mathbf{\sigma}\cdot m_{f}\mathbf{v}_{r}(t_{\mes})\chi_{r}
\right]\left[\bar{\chi}_{r}\mathbf{\sigma}\cdot m_{f}\mathbf{v}_{r}(t_{\mic})\chi_{i}+\bar{\chi}_{r}
\mathbf{\sigma}\cdot(m_{f}\mathbf{v}_{i}(t_{\mic})-\mathbf{q})\chi_{i}\right]\\
& \times\rho_{jr'}(\mathbf{v})e^{-\mathrm{i}(P^{0}-k^{0}+q^{0})(t_{\mes}-t_{\mic})}\\
& +\left[\bar{\chi}_{j}\mathbf{\sigma}\cdot m_{f}\mathbf{v}_{j}(t_{\mes})\chi_{r}+\bar{\chi}_{j}
\mathbf{\sigma}\cdot(m_{f}\mathbf{v}_{r}(t_{\mes})+\mathbf{q})\chi_{r}
\right]\left[\bar{\chi}_{r'}\mathbf{\sigma}\cdot(m_{f}\mathbf{v}_{r'}(t_{\mic})+\mathbf{q})\chi_{i}
+\bar{\chi}_{r'}\mathbf{\sigma}\cdot m_{f}\mathbf{v}_{i}(t_{\mic})\chi_{i}\right]\\
& \times\rho_{rr'}(\mathbf{v})e^{-\mathrm{i}(k^{0}-P'^{0}+q^{0})(t_{\mes}-t_{\mic})}\\
& -\left[\bar{\chi}_{r'}\mathbf{\sigma}\cdot m_{f}\mathbf{v}_{r'}(t_{\mes})\chi_{r}+\bar{\chi}_{r'}
\mathbf{\sigma}\cdot(m_{f}\mathbf{v}_{r}(t_{\mes})+\mathbf{q})\chi_{r}
\right]\left[\bar{\chi}_{j}\mathbf{\sigma}\cdot(m_{f}\mathbf{v}_{j}(t_{\mic})+\mathbf{q})\chi_{r'}
+\bar{\chi}_{j}\mathbf{\sigma}\cdot m_{f}\mathbf{v}_{r'}(t_{\mic})\chi_{r'}\right]\\
& \times\rho_{ri}(\mathbf{v})e^{-\mathrm{i}(k^{0}-P'^{0}+q^{0})(t_{\mes}-t_{\mic})}\\
& +\left[\bar{\chi}_{j}\mathbf{\sigma}\cdot(m_{f}\mathbf{v}_{j}(t_{\mes})-\mathbf{q})\chi_{r}
+\bar{\chi}_{j}\mathbf{\sigma}\cdot m_{f}\mathbf{v}_{r}(t_{\mes})\chi_{r}
\right]\left[\bar{\chi}_{r'}\mathbf{\sigma}\cdot m_{f}\mathbf{v}_{r'}(t_{\mic})\chi_{i}+
\bar{\chi}_{r'}\mathbf{\sigma}\cdot(m_{f}\mathbf{v}_{i}(t_{\mic})-\mathbf{q})\chi_{i}\right]\\
& \times\rho_{rr'}(\mathbf{v})e^{-\mathrm{i}(P^{0}-k^{0}+q^{0})(t_{\mes}-t_{\mic})}\Big\}\,,
\end{split}
\end{equation}
where $P^{0}=E(\mathbf{k}-\mathbf{q})$ and $P'^{0}=E(\mathbf{k}+\mathbf{q})$ and also, we employed the approximation $\rho(\mathbf{k}-\mathbf{q})\approx\rho(\mathbf{k}+\mathbf{q})\approx\rho(\mathbf{k})$ as well as $q^{0}\approx m_{a}$. Also, we utilized the following abbreviation
\begin{equation}\label{dpqpp}
\frac{d^{3}\mathbf{v}_{a}}{(2\pi)^{3}}\equiv\mathrm{D}\mathbf{v}_{a}\,.
\end{equation}

Now, for the right-hand side of Eq. \eqref{qbe3} one should first take the time integral over $t_{\mic}$. Then, by converting the exponential functions gained to the sinc functions and assuming energy conservation, one can calculate the integration over $\mathbf{v}_{a}$ in Eq. \eqref{qbe3}.

\section{Estimate of the induced phase}\label{eot}

We numerically solve the QBE for both the forward and the collision terms. The forward scattering term induces only the relative phase shift, while the collision term is responsible for both the relative phase shift and decoherence. Our numerical investigation shows that the effect of the collision term, however, is dominated by the forward scattering term. Accordingly, the induced decoherence is suppressed because of the smallness of the collision term effect. In the following, we will discuss these results in more detail.

Using Eqs. \eqref{zetai}, \eqref{qbefs3}, and \eqref{qbe3}, the following final results for the time evolution of the Bloch vector components associated with both the forward scattering and collision terms are obtained as the following three coupled differential equations
\begin{equation}\label{zetaiev}
\begin{split}
& \dot{\zeta}_{1}(\mathbf{v},t_{\mes})=Agm_{a}\sqrt{2\pi}e^{-\mathrm{i}m_{a}t_{\mes}}
\left\{v_{3}\zeta_{1}(t_{\mes})\left(1-2
\Theta\left[t_{\mes}-\frac{T}{2}\right]\right)-\mathrm{i}v_{2}\zeta_{3}(t_{\mes})+v_{1}\right\}\\
& \hspace{1.6 cm}-\frac{g^{2}\pi\rho}{16m_{f}^{2}}\left\{t_{\mes}\left(m_{a}^{2}
W_{1}-2m_{a}m_{f}W_{2}+4m_{f}^{2}W_{3}\right)-2m_{f}v_{3}W_{4}\Theta\left[t_{\mes}-\frac{T}{2}\right]
\right\}\,,\\
& \dot{\zeta}_{2}(\mathbf{v},t_{\mes})=Agm_{a}\sqrt{2\pi}e^{-\mathrm{i}m_{a}t_{\mes}}
\left\{v_{3}\zeta_{2}(t_{\mes})\left(1-2
\Theta\left[t_{\mes}-\frac{T}{2}\right]\right)+\mathrm{i}v_{1}\zeta_{3}(t_{\mes})+v_{2}\right\}\\
& \hspace{1.6 cm}+\frac{g^{2}\pi\rho}{16m_{f}^{2}}\left\{t_{\mes}\left(m_{a}^{2}
Z_{1}-2m_{a}m_{f}Z_{2}+4m_{f}^{2}Z_{3}\right)+2m_{f}v_{3}Z_{4}\Theta\left[t_{\mes}-\frac{T}{2}\right]
\right\}\,,\\
& \dot{\zeta}_{3}(\mathbf{v},t_{\mes})=Agm_{a}\sqrt{2\pi}e^{-\mathrm{i}m_{a}t_{\mes}}
\left\{v_{3}\zeta_{3}(t_{\mes})\left(1-2
\Theta\left[t_{\mes}-\frac{T}{2}\right]\right)-\mathrm{i}v_{1}\,\zeta_{2}(t_{\mes})
+\mathrm{i}v_{2}\zeta_{1}(t_{\mes})\right\}\\
& \hspace{1.6 cm}-\frac{g^{2}\pi\rho}{16m_{f}^{2}}\left\{t_{\mes}\left(m_{a}^{2}
Y_{1}-2m_{a}m_{f}Y_{2}+4m_{f}^{2}Y_{3}\right)+2m_{f}v_{3}Y_{4}\Theta\left[t_{\mes}-\frac{T}{2}\right]
\right\}\,,
\end{split}
\end{equation}
where the explicit forms of coefficients $W_{i}, Z_{i}$, and $Y_{i}$ are given in the Appendix \ref{ap2}.

We provide the system such that the initial phase is $\varphi_{0}=0$. Then, from Eq. \eqref{decm}, one can obtain the relative phase shift as
\begin{equation}\label{phsh}
\Delta\varphi=\varphi-\varphi_{0}=\left|\tan^{-1}\left[\frac{\Im(\zeta_{1}(t_{\mes}))-\Re(\zeta_{2}
(t_{\mes}))}{\Im(\zeta_{2}(t_{\mes}))+\Re(\zeta_{1}(t_{\mes}))}\right]\right|\,.
\end{equation}
This equation gives us the time evolution of phase measure within the decoherence factor \eqref{decm} during the process. One can numerically solve the set of Eqs. \eqref{zetaiev} to calculate the relative phase shift \eqref{phsh} due to the interaction of neutral atoms with ALPs. To this end, we consider two types of neutral atoms, $^{3}$He and $^{87}$Rb, moving in the SG interferometer.

Next, we turn our attention to the phase sensitivity of the system and the time duration of the process. The phase sensitivity is defined as the smallest measurable phase shift, which is determined by the number of nucleons, typical contrast, and other factors. On the other hand, the time duration of the process is the total time during which the effective spin $1/2$ atom are within the SG interferometer. Among the available values chosen by the proposed experiments, we assume that the minimum phase (phase sensitivity) provided by the setup is of the order of $\Delta\varphi_{min}=10^{-10}\,\mathrm{Rad}$ and the time duration of the process is of the order of $T=2.6\times 10^{-4}\,\mathrm{s}$ \cite{D2022}.

\begin{figure}[htb]
\centering
\includegraphics[width=0.90\columnwidth]{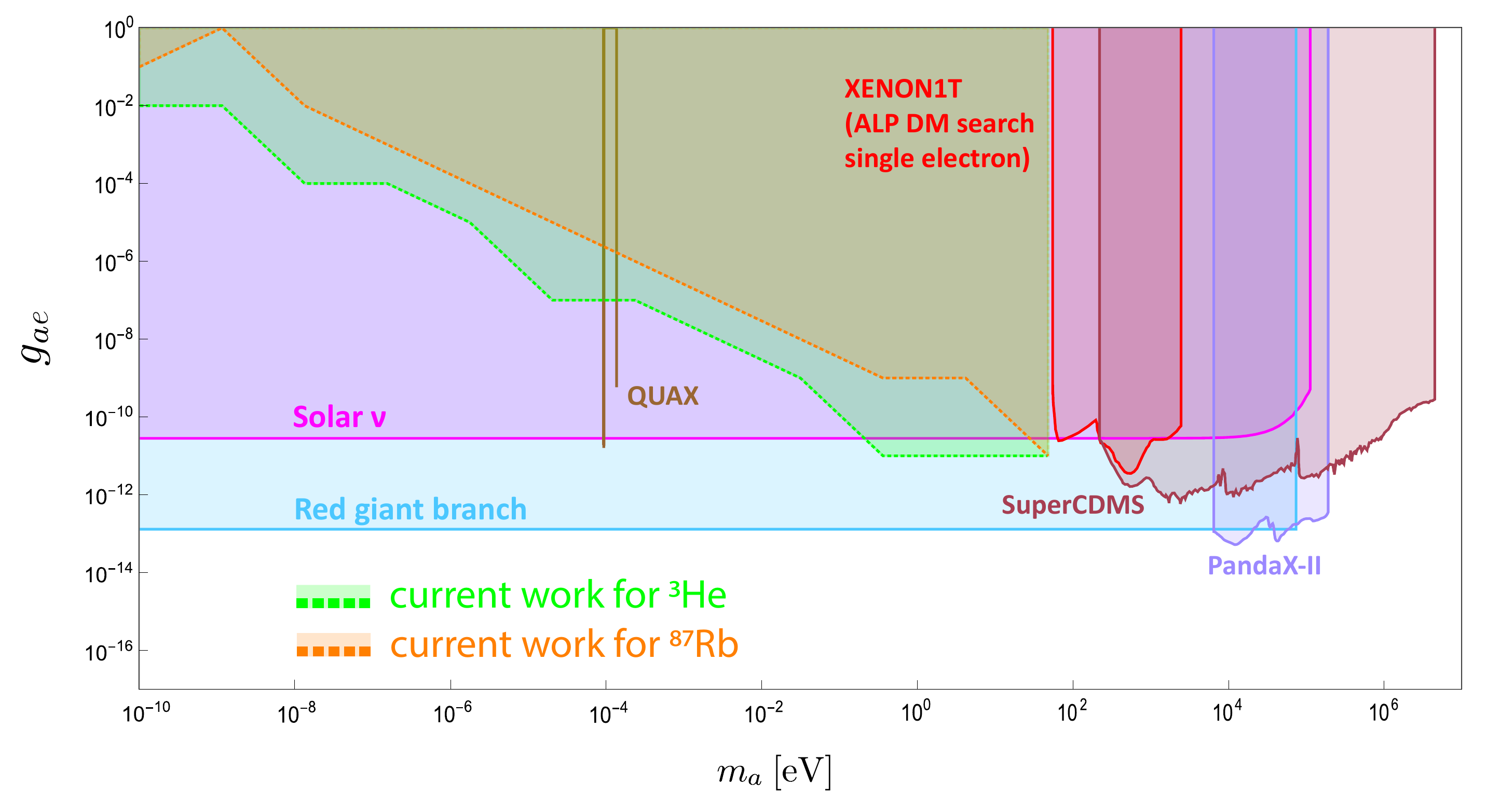}
\caption{\label{Fig3}\small{\emph{The exclusion areas in the plane of ALPs mass and the coupling constant of ALPs-electron interaction for $^{\textit{3}}$He and $^{\textit{87}}$Rb compared with other proposals \cite{Hare2020}.}}}
\end{figure}

We plot the regions that our scheme can exclude for these particles. We compare the data from Underground Detectors: PandaX-II \cite{Fu2017}, SuperCDMS \cite{Aralis2020}, XENON1T (ALP DM search single electron) \cite{Aprile2022}; data from Haloscopes: QUAX \cite{Crescini2018,Crescini2020}; and Astro bound: Red giant branch \cite{Capozzi2020} and Solar $\nu$ \cite{Gondolo2009}, with total outcome of $^{3}$He and $^{87}$Rb, in Fig. \eqref{Fig3}. In this figure, $g_{ae}$ (equivalent to $g_{af}$ in Eq. \eqref{ccoi}) is the dimensionless coupling constant of ALP-electron interaction, which is based on the assumption that ALPs interact with the electrons of the atom that we have considered as a two-level system. As Fig. \eqref{Fig3} illustrates, our scheme has excluded the area between $10^{-10}\leq m_{a}\leq 10^{2}\,\mathrm{eV}$ and $10^{-13}\leq g_{ae}\leq 10^{0}$. Again, it is worth noting that the effect of the collision term is tiny, so its corresponding induced decoherence is negligible.

\section{Conclusion and discussion}\label{cad}

This paper presented a new scheme for the detection of ALPs as DM by the SG interferometer. ALPs interact with neutral atoms moving in the SG interferometer. The neutral atoms are considered to be in a superposition state, and the time evolution of their corresponding density matrix is derived by considering the forward scattering and the collision terms of the QBE. The interaction results in a relative phase shift between the initial and final superposition states of the effectively two-level neutral atoms. We considered two neutral atoms, $^{3}$He and $^{87}$Rb, moving in the SG interferometer. Taking a zero initial phase into account, we calculated the induced relative phase shift. This way, we plot the exclusion areas for each neutral atom compared to other proposals. Consequently, we found that our scheme has excluded the area between $10^{-10}\leq m_{a}\leq 10^{2}\,\mathrm{eV}$ and $10^{-13}\leq g_{ae}\leq 10^{0}$. We also propose considering other schemes that use lighter neutral particles such as neutrons with the same minimum phase. Providing such conditions can exclude a wider region compared with other experiments.

\begin{acknowledgments}

MH acknowledges fruitful support by Iran's National Elites Foundation under the AMS program, and is grateful for the financial support from the Isfahan University of Technology. The authors would like to thank M. Abdi for his insightful discussions and comments. The authors also appreciate the respectful referee for carefully reading the original manuscript and their insightful comments, which have improved the quality of the paper.

\end{acknowledgments}

\appendix

\section{On inaccuracy of the full-loop SG interferometer}\label{ap1}

In this study, we utilize the SG interferometer to design an experimental scheme for detecting ALPs as a favored candidate for DM. In an SG interferometer, an inhomogeneous magnetic field rather than laser pulses are responsible for producing spatial superposition in wave packets of the effective spin $1/2$ atoms over relatively large spatial separation. Four magnetic field gradients are utilized to split, stop, accelerate back, and stop the wave packets to complete a full-loop in the SG interferometer. The spin state of the recombined wave packets is used as an interference signal by the full-loop SG interferometer, which recombines the wave packets in both position and momentum. For the purposes of the current study, since the effectively spin $1/2$ atoms can absorb or even scatter the laser pulses, one cannot utilize the usual laser-based interferometers (see more details in Ref. \cite{Henkel2022}). Realistic magnetic field gradients, however, have some inaccuracy, for example, temporal fluctuations, etc.

A formidable challenge known as the Humpty-Dumpty effect \cite{Englert1988,Schwinger1988,Scully1988,Oliveira2006} makes the ideal full-loop SG interferometer impractical because it leads to decoherence in the output. The Humpty-Dumpty effect briefly states that the interferometer demands a ``fantastic accuracy'' \cite{Bohm1951} of the magnetic field gradients to preserve the coherence between superposed wave packets propagating in the different arms of the interferometer, leading to a full-loop SG interferometer. In this scene, if the magnetic field gradients are ideally controlled, a significant amount of coherence can then be recovered \cite{Englert1988}, and the interferometer is approximately full-loop, while a quantum mechanical point of view requiring realistic magnetic field gradients results in decoherence or loss of coherence, which prevents the SG interferometer from being full-loop \cite{Schwinger1988}.

To achieve a sufficiently high full-loop SG interferometer, the time-reversal operation of the splitting process by the first two magnetic gradients should be precisely the same as the recombination process operated by the final two magnetic gradients. This demands stable and accurate operations on the effective spin $1/2$ atoms within the interferometer. Therefore, one should adequately minimize the final relative distance and momentum between the wave packets at the final stage. Additionally, the time duration of the splitting process and recombination process should be exactly equal. Furthermore, the magnetic field gradients should be strong enough and quickly switchable to significantly reduce the time duration of the whole process $T$. Moreover, the location and structure of the magnetic field gradients play a crucial role and should be precise. In addition, one should consider the ultracold wave packets of the effectively spin $1/2$ atoms with minimal uncertainty rather than the thermal beams of these atoms. Taking into account the abovementioned considerations, we find that the output of the full-loop SG interferometer is highly coherent.

In addition to the Humpty-Dumpty effect, which prevents the loop of the SG interferometer from being closed, another source of decoherence in the SG interferometer is the coupling of the effectively spin $1/2$ ultracold atoms to the environment (for more details, see Ref. \cite{Henkel2022} and references therein). To suppress the coupling of spin $1/2$ ultracold atoms with the environment, the time duration of the whole process should be much shorter than the decoherence time arising from such a coupling \cite{Oliveira2006}. We also assume that our experimental scheme is operated in vacuum conditions. Moreover, such a coupling can be the interaction of the effective spin $1/2$ ultracold atoms with an electromagnetic field and the bremsstrahlung effect \cite{Pratt1985} (which exists even in the vacuum), as it is well known that it can cause decoherence \cite{Breuer2002,Breuer2001}. It can be shown in Ref. \cite{Manshouri2022} that despite the bremsstrahlung effect leading to decoherence, its impact on dephasing time is less than the ALP-atom interaction. Investigations from Ref. \cite{Manshouri2022} demonstrate that the induced phase shift due to the bremsstrahlung effect is of the order of $\delta\phi\sim 10^{-27}\,\mathrm{Rad}$ and $10^{-28}\,\mathrm{Rad}$ for $^{3}$He and $^{87}$Rb, respectively, which is in agreement with Refs. \cite{Breuer2002,Breuer2001}. Therefore, in this study, the decoherence induced by coupling to the environment and the bremsstrahlung effect can be entirely neglected.

\section{Coefficients of the collision term}\label{ap2}

To calculate the relative phase shift, the coefficients introduced in Eq. \eqref{zetaiev} were lengthy. We found them through a straightforward calculation. In the following, we provide their explicit form 
\begin{equation}\label{defi1}
\begin{split}
W_{1} & =8\zeta_{1}(t_{\mes})\sigma_{v}^{2}+V^{E}_{1}\Big(\big(\zeta_{1}(t_{\mes})-1\big)V^{E}_{1}-\Big(\zeta_{2}
(t_{\mes})\big(2V^{E}_{2}+\mathrm{i}V^{E}_{3}\big)-\mathrm{i}\zeta_{3}(t_{\mes})\big(2V^{E}_{2}+3\mathrm{i}
V^{E}_{3}\big)\Big)\Big)\\
& +\big(V^{E}_{2}+\mathrm{i}V^{E}_{3}\big)\Big(\zeta_{1}(t_{\mes})\big(3V^{E}_{2}-
4\mathrm{i}V^{E}_{3}\big)+V^{E}_{2}\Big)\,,
\end{split}
\end{equation}
\begin{equation}\label{defi2}
\begin{split}
W_{2} & =v_{2}\Big(2\big(\zeta_{2}(t_{\mes})+\mathrm{i}\zeta_{3}(t_{\mes})\big)V^{E}_{1}-\big(2V^{E}_{2}+
\mathrm{i}V^{E}_{3}\big)\big(\zeta_{1}(t_{\mes})-1\big)\Big)+v_{1}\Big(2\big(\zeta_{1}(t_{\mes})-1\big)V^{E}_{1}
+\big(2V^{E}_{2}-\mathrm{i}V^{E}_{3}\big)\\
& \times\big(\zeta_{2}(t_{\mes})+\mathrm{i}\zeta_{3}(t_{\mes})\big)\Big)-
\Big(\big(\zeta_{1}(t_{\mes})-1\big)V^{E}_{1}-\big(\zeta_{2}(t_{\mes})+5\mathrm{i}\zeta_{3}(t_{\mes})\big)
V^{E}_{2}+4\mathrm{i}\zeta_{2}(t_{\mes})V^{E}_{3}\Big)v_{3}\,,
\end{split}
\end{equation}
\begin{equation}\label{defi3}
\begin{split}
W_{3} & =v_{1}\Big(\big(\zeta_{1}(t_{\mes})-1\big)\big(v_{1}-v_{3}\big)-2\big(\zeta_{2}(t_{\mes})
+\mathrm{i}\zeta_{3}(t_{\mes})\big)v_{2}\Big)+v_{2}\Big(\big(\zeta_{2}(t_{\mes})+\mathrm{i}\zeta_{3}
(t_{\mes})\big)v_{3}+\big(3\zeta_{2}(t_{\mes})\\
& -\mathrm{i}\zeta_{3}(t_{\mes})\big)v_{2}\Big)\,,
\end{split}
\end{equation}
\begin{equation}\label{defi4}
\begin{split}
W_{4} & =m_{a}\Big(2\Big(\big(\zeta_{1}(t_{\mes})-1\big)V^{E}_{1}-\big(\zeta_{2}(t_{\mes})+5\mathrm{i}\zeta_{3}
(t_{\mes})\big)V^{E}_{2}+4\mathrm{i}\zeta_{2}(t_{\mes})V^{E}_{3}\Big)t_{\mes}+\Big(-\big(\zeta_{1}(t_{\mes})
-1\big)V^{E}_{1}\\
& +\zeta_{2}(t_{\mes})\big(V^{E}_{2}-2\mathrm{i}V^{E}_{3}\big)+3\mathrm{i}\zeta_{3}(t_{\mes})
V^{E}_{2}\Big)T\Big)-4m_{f}\Big(\Big(t_{\mes}-\frac{T}{2}\Big)\Big(\big(\zeta_{1}(t_{\mes})-1\big)v_{1}-
\zeta_{2}(t_{\mes})v_{2}\Big)\\
& -\mathrm{i}\zeta_{3}(t_{\mes})\Big(\frac{T}{2}+t_{\mes}\Big)v_{2}\Big)\,,
\end{split}
\end{equation}
\begin{equation}\label{defi5}
\begin{split}
Z_{1} & =V^{E}_{1}\Big(\big(3\zeta_{2}(t_{\mes})-\mathrm{i}\zeta_{3}(t_{\mes})\big)V^{E}_{1}+\mathrm{i}
\big(\zeta_{1}(t_{\mes})-1\big)V^{E}_{3}\Big)+V^{E}_{2}\Big(\big(\zeta_{2}(t_{\mes})+\mathrm{i}\zeta_{3}
(t_{\mes})\big)V^{E}_{2}-2\big(\zeta_{1}(t_{\mes})+1\big)V^{E}_{1}\\
& -\mathrm{i}\big(\zeta_{2}(t_{\mes})-3\mathrm{i}\zeta_{3}(t_{\mes})\big)V^{E}_{3}\Big)+4\zeta_{2}
(t_{\mes})\big(2\sigma_{v}^{2}+{V^{E}_{3}}^{2}\big)\,,
\end{split}
\end{equation}
\begin{equation}\label{defi6}
\begin{split}
Z_{2} & =\big(2V^{E}_{2}+\mathrm{i}V^{E}_{3}\big)\big(\zeta_{1}(t_{\mes})-1\big)v_{1}+2V^{E}_{1}\Big(\big(\zeta_{1}
(t_{\mes})-1\big)v_{2}-\big(\zeta_{2}(t_{\mes})+\mathrm{i}\zeta_{3}(t_{\mes})\big)v_{1}\Big)+\big(2V^{E}_{2}-
\mathrm{i}V^{E}_{3}\big)\big(\zeta_{2}(t_{\mes})\\
& +\mathrm{i}\zeta_{3}(t_{\mes})\big)v_{2}-\Big(\big(\zeta_{1}(t_{\mes})-1\big)V^{E}_{2}+\big(\zeta_{2}
(t_{\mes})+5\mathrm{i}\zeta_{3}(t_{\mes})\big)V^{E}_{1}-4\mathrm{i}\zeta_{1}(t_{\mes})V^{E}_{3}\Big)v_{3}\,,
\end{split}
\end{equation}
\begin{equation}\label{defi7}
\begin{split}
Z_{3}=v_{2}\Big(\big(\zeta_{1}(t_{\mes})-1\big)v_{3}+\big(\zeta_{2}(t_{\mes})+\mathrm{i}\zeta_{3}(t_{\mes})\big)
v_{2}\Big)+v_{1}\Big(-v_{3}+\big(3\zeta_{2}(t_{\mes})-\mathrm{i}\zeta_{3}(t_{\mes})\big)v_{1}-2\big(\zeta_{1}
(t_{\mes})+1\big)v_{2}\Big)\,,
\end{split}
\end{equation}
\begin{equation}\label{defi8}
\begin{split}
Z_{4} & =m_{a}\Big(\Big(\big(\zeta_{1}(t_{\mes})-1\big)V^{E}_{2}+\big(\zeta_{2}(t_{\mes})+3\mathrm{i}\zeta_{3}
(t_{\mes})\big)V^{E}_{1}-2\mathrm{i}\zeta_{1}(t_{\mes})V^{E}_{3}\Big)T-2\Big(\big(\zeta_{1}(t_{\mes})-1\big)
V^{E}_{2}+\big(\zeta_{2}(t_{\mes})\\
& +5\mathrm{i}\zeta_{3}(t_{\mes})\big)V^{E}_{1}-4\mathrm{i}\zeta_{1}(t_{\mes})
V^{E}_{3}\Big)t_{\mes}\Big)+4m_{f}\Big(\Big(t_{\mes}-\frac{T}{2}\Big)\Big(\big(\zeta_{1}(t_{\mes})-1\big)v_{2}
+\zeta_{2}(t_{\mes})v_{1}\Big)\\
& +\mathrm{i}\zeta_{3}(t_{\mes})\Big(\frac{T}{2}+t_{\mes}\Big)v_{1}\Big)\,,
\end{split}
\end{equation}
\begin{equation}\label{defi9}
\begin{split}
Y_{1}=\mathrm{i}\Big(\big(\zeta_{2}(t_{\mes})-3\mathrm{i}\zeta_{3}(t_{\mes})\big)\big(2\sigma_{v}^{2}
+{V^{E}_{1}}^{2}+{V^{E}_{2}}^{2}\big)+\mathrm{i}V^{E}_{3}\Big(\big(3\zeta_{2}(t_{\mes})-\mathrm{i}
\zeta_{3}(t_{\mes})\big)V^{E}_{2}+\big(1+3\zeta_{1}(t_{\mes})\big)V^{E}_{1}\Big)\Big)\,,
\end{split}
\end{equation}
\begin{equation}\label{defi10}
\begin{split}
Y_{2} & =\mathrm{i}\Big(v_{1}\Big(2\big(\zeta_{2}(t_{\mes})+\mathrm{i}\zeta_{3}(t_{\mes})\big)V^{E}_{1}
-\mathrm{i}\big(\zeta_{1}(t_{\mes})-1\big)V^{E}_{3}\Big)+\big(2V^{E}_{2}-\mathrm{i}V^{E}_{3}\big)\big(\zeta_{2}
(t_{\mes})+\mathrm{i}\zeta_{3}(t_{\mes})\big)v_{2}\\
& +v_{3}\Big(\big(5\zeta_{2}(t_{\mes})+\mathrm{i}\zeta_{3}(t_{\mes})\big)V^{E}_{1}+V^{E}_{2}\Big)\Big)\,,
\end{split}
\end{equation}
\begin{equation}\label{defi11}
\begin{split}
Y_{3}=\mathrm{i}\Big(v_{2}\Big(\big(\zeta_{2}(t_{\mes})-3\mathrm{i}\zeta_{3}(t_{\mes})\big)v_{2}-\big(\zeta_{1}
(t_{\mes})-1\big)v_{3}\Big)+v_{1}\Big(\zeta_{2}(t_{\mes})\big(v_{1}+v_{3}\big)+\mathrm{i}\zeta_{3}(t_{\mes})
\big(v_{3}-3v_{1}\big)\Big)\Big)\,,
\end{split}
\end{equation}
\begin{equation}\label{defi12}
\begin{split}
Y_{4} & =\mathrm{i}\Big(2m_{a}t_{\mes}\Big(\big(1-5\zeta_{1}(t_{\mes})\big)V^{E}_{2}+\big(5\zeta_{2}(t_{\mes})
+\mathrm{i}\zeta_{3}(t_{\mes})\big)V^{E}_{1}\Big)-m_{a}T\Big(\big(1+3\zeta_{1}(t_{\mes})\big)V^{E}_{2}+
\big(3\zeta_{2}(t_{\mes})\\
& +\mathrm{i}\zeta_{3}(t_{\mes})\big)V^{E}_{1}\Big)-4m_{f}t_{\mes}\Big(\big(\zeta_{2}
(t_{\mes})+\mathrm{i}\zeta_{3}(t_{\mes})\big)v_{1}-\big(\zeta_{1}(t_{\mes})-1\big)v_{2}\Big)
+2m_{f}T\Big(\big(\zeta_{1}(t_{\mes})+1\big)v_{2}\\
& -\big(\zeta_{2}(t_{\mes})+\mathrm{i}\zeta_{3}(t_{\mes})\big)v_{1}\Big)\Big)\,.
\end{split}
\end{equation}

\end{document}